\documentclass{ws-procs9x6}   
\usepackage{makecell}         
\begin{document}
\title{Hyperon resonances and meson-baryon interactions in isospin 1}

\author{K. P. Khemchandani$^*$}

\address{Departamento de f\'isica, Universidade Federal de S\~ao Paulo, Campus Diadema, 09913-030, S\~ao Paulo, Brazil..\\
$^*$E-mail: kanchan.khemchandani@unifesp.br}

\author{A. Mart\'inez Torres}

\address{Instituto de F\'isica, Universidade de S\~ao Paulo, C.P 66318, 05314-970 S\~ao Paulo, SP, Brazil.\\
E-mail: amartine@if.usp.br}

\author{J. A. Oller}
\address{Departamento de F\`isica. Universidad de Murcia. E-30071 Murcia. Spain.\\
E-mail: oller@um.es}

\begin{abstract}
In this work we extend our formalism to study meson-baryon interactions by including $s$- and $u$-channel diagrams for pseudoscalar-baryon systems. We study the coupled systems with strangeness $-1$ and focus on studying the isospin-1 resonance(s), especially in the energy region around 1400 MeV.  By constraining the  model parameters  to fit the cross section data available on several processes involving relevant channels, we find resonances in the isoscalar as well as the  isovector sector in the energy region around 1400 MeV.
\end{abstract}


\bodymatter

\section{Introduction}\label{aba:sec1}
There exists a large consensus on the nature of the first excited state of the isoscalar strange baryon, $\Lambda(1405)$, as being a moleculelike state rather than a standard three valence quark baryon with one of the quarks in a higher angular momentum state (see Refs.~\cite{Jido:2003cb,Oller:2006jw,Menadue:2011pd,Mai:2012dt} for some of the most-cited articles on the topic). In the isovector case, on the other hand, the lightest  $\Sigma$ with the same spin-parity as $\Lambda(1405)$, $1/2^-$, listed by the particle data group \cite{pdg}, is $\Sigma(1620)$, which possesses  a poor status of information on its properties. Various studies \cite{Wu:2009tu,Wu:2009nw,Gao:2010hy,Xie:2014zga,Xie:2017xwx,Oller:2000fj,Guo,Moriya:2013eb}, though, seem to indicate that the existence of a $1/2^-$ $\Sigma$, with a much lower mass ($\sim$ 1400 MeV),  is needed to describe the experimental data on the processes like: $K^- p \to \Lambda \pi^+ \pi^-$, $\gamma N \to  K^+ \pi \Lambda$, $\Lambda p \to \Lambda p \pi^0$, $\Lambda^+_c \to \eta \pi^+ \Lambda$, $\gamma + p \to K^+ + \Sigma^{\pm,0} + \pi^{\mp,0}$. In some of these former works the existence of a $\Sigma$ state with mass around 1380 MeV is discussed and in some others a scenario with two close lying poles is discussed. In such a situation, we find it useful to study the hyperon resonances with isospin 1 in the energy region around 1400 MeV. Indeed, in our previous work \cite{Khemchandani:2012ur} based on a coupled channel formalism, including vector and pseudoscalar mesons, we found two $\Sigma$ resonances: $1427 - i145$ MeV, $1438 - i198$ MeV, which are much wider than those claimed in Refs.~\cite{Wu:2009tu,Wu:2009nw,Gao:2010hy,Xie:2014zga,Xie:2017xwx,Oller:2000fj,Guo,Moriya:2013eb}. In Ref.~\cite{Khemchandani:2012ur}, though, we considered a more elaborate formalism to obtain the lowest order vector meson-baryon (VB) interaction by calculating the $s$-, $t$-, $u$- channel diagrams and a contact interaction, while relying on the  Weinberg-Tomozawa term alone for the pseudoscalar-baryon (PB)  interaction. Inspired by the works in Refs.~\cite{Oller:2000fj,Guo}, we now include $s$- and $u$-channel diagrams for the pseudoscalar interaction and solve the Bethe-Salpeter equation.  Further,  we constrain the parameters by fitting the amplitudes to the cross section data on the processes: $K^- p \to K^- p,~\bar K^0 n,~ \eta \Lambda,~\pi^0 \Lambda~\pi^0 \Sigma^0,~ \pi^\pm \Sigma^\mp$, on the energy level shift and width of the $1s$ state of the kaonic hydrogen from Ref.~\cite{Bazzi:2011zj} and certain cross section ratios near the threshold (more details can be found in the article~\cite{Khemchandani:2018amu}).

\section{Formalism and results}
The  amplitudes (for the contact interaction, $s$- and $u$-channel diagrams), for the pseudoscalar meson-baryon interactions, are obtained from the lowest order chiral Lagrangian  
\begin{align}
&\mathcal{L}_{PB} = \langle \bar B i \gamma^\mu \partial_\mu B  + \bar B i \gamma^\mu[ \Gamma_\mu, B] \rangle - M_{B} \langle \bar B B \rangle
\\\nonumber&+  \frac{1}{2} D^\prime \langle \bar B \gamma^\mu \gamma_5 \{ u_\mu, B \} \rangle + \frac{1}{2} F^\prime \langle \bar B \gamma^\mu \gamma_5 [ u_\mu, B ] \rangle,\label{LPB}
\end{align}
where  $u_\mu = i u^\dagger \partial_\mu U u^\dagger$, and
\begin{eqnarray}
\Gamma_\mu &=& \frac{1}{2} \left( u^\dagger \partial_\mu u + u \partial_\mu u^\dagger  \right), 
\, U=u^2 = {\textrm exp} \left(i \frac{P}{f_P}\right),\label{gammau}
\end{eqnarray}
with $P$ and $B$ representing the matrices of the pseudoscalar mesons and the octet baryons.
The $s$-, $t$-, $u$-amplitudes for the vector meson-baryon interactions are obtained from the Lagrangian~\cite{vbvb}
\begin{eqnarray} \label{vbb}
&\mathcal{L}_{\textrm VB}& = -g \Biggl\{ \langle \bar{B} \gamma_\mu \left[ V_8^\mu, B \right] \rangle\!+\! \langle \bar{B} \gamma_\mu B \rangle  \langle  V_8^\mu \rangle  
\Biggr. + \frac{1}{4 M}\! \left( F \langle \bar{B} \sigma_{\mu\nu} \left[ V_8^{\mu\nu}, B \right] \rangle \! \right.\\\nonumber
&&\left.+\! D \langle \bar{B} \sigma_{\mu\nu} \left\{V_8^{\mu\nu}, B \right\} \rangle\right) +  \Biggl.  \langle \bar{B} \gamma_\mu B \rangle  \langle  V_0^\mu \rangle  
+ \frac{ C_0}{4 M}  \langle \bar{B} \sigma_{\mu\nu}  V_0^{\mu\nu} B  \rangle  \Biggr\},
\end{eqnarray}
where $V$ refers to the matrix of vector mesons, with the numbers 0/8 in the subscript indicating the singlet/octet part of the wave function (arising from the consideration of $\omega-\phi$ mixing).
The contribution of the  $V_{\mu\nu}$ tensor gives rise to yet another diagram, a contact interaction. 

The transition amplitudes between the two types of meson-baryons systems are obtained from the Lagrangian~\cite{pbvb}
\begin{eqnarray} \label{pbvbeq}
\mathcal{L}_{\rm PBVB}\! =\! \frac{-i g_{PBVB}}{2 f_v} \left( F^\prime \langle \bar{B} \gamma_\mu \gamma_5 \!\left[ \left[ P, V^\mu \right], B \right]\, \rangle\!+\!
D^\prime \langle \bar{B} \gamma_\mu \gamma_5 \left\{ \left[ P, V^\mu \right], B \!\right\}  \!\rangle\! \right)\!.
\end{eqnarray}

With all the amplitudes  determined from the above Lagrangians, we solve the Bethe-Salpeter equation in the on-shell factorization form for the coupled channels: $\bar KN$, $K \Xi$, $\pi \Sigma$, $\eta \Lambda$, $\pi \Lambda$, $\eta \Sigma$, $\bar K^*N$, $K^* \Xi$, $\rho \Sigma$, $\omega \Lambda$, $\phi \Lambda$, $\rho \Lambda$, $\omega \Sigma$, $\phi \Sigma$.  The detailed amplitudes for the different processes are given in Ref.~\cite{Khemchandani:2018amu}. To solve the scattering equations, we need to regularize the divergent loop functions, which is done by using the dimensional regularizations method. We, thus, have a parameter at hand, which is the subtraction constant needed for the calculation of the loop function. We treat the subtraction constant for each channel as a free parameter, which is fitted to reproduce the experimental data mentioned in the previous section.  In addition to the subtraction constants, we treat the coupling between the PB and VB channels, and the decay constants for the two types of mesons as a parameter to be fixed by the data. When fitting the data, we calculate the $\chi^2$ per degree of freedom as 
\begin{align}
\chi^2_\text{d.o.f}=\frac{\sum\limits_{k=1}^N n_k}{N (\sum\limits_{k=1}^N n_k-n_p)}\sum\limits_{k=1}^N \frac{\chi^2_k}{n_k},
\end{align}
where $N$ is the number of different  sets of data, $n_k$ represents the number of data points in the $k$th data set, $n_p$ is the number of free parameters, and the $\chi^2$ for the $k$th data set is obtained as
\begin{align}
\chi^2_k=\sum\limits_{i=1}^{n_k}\frac{(y^{\text{th}}_{k;i}-y^{\text{exp}}_{k;i})^2}{\sigma^2_{k;i}},
\end{align}
with $y^{\text{exp}}_{k;i}$ ($y^{\text{th}}_{k;i}$) representing the $i$th experimental (theoretical) data of the $k$th  set and $\sigma^2_{k;i}$ the related standard deviation.

\begin{figure}[h!]
\begin{center}
\includegraphics[width=2.5in]{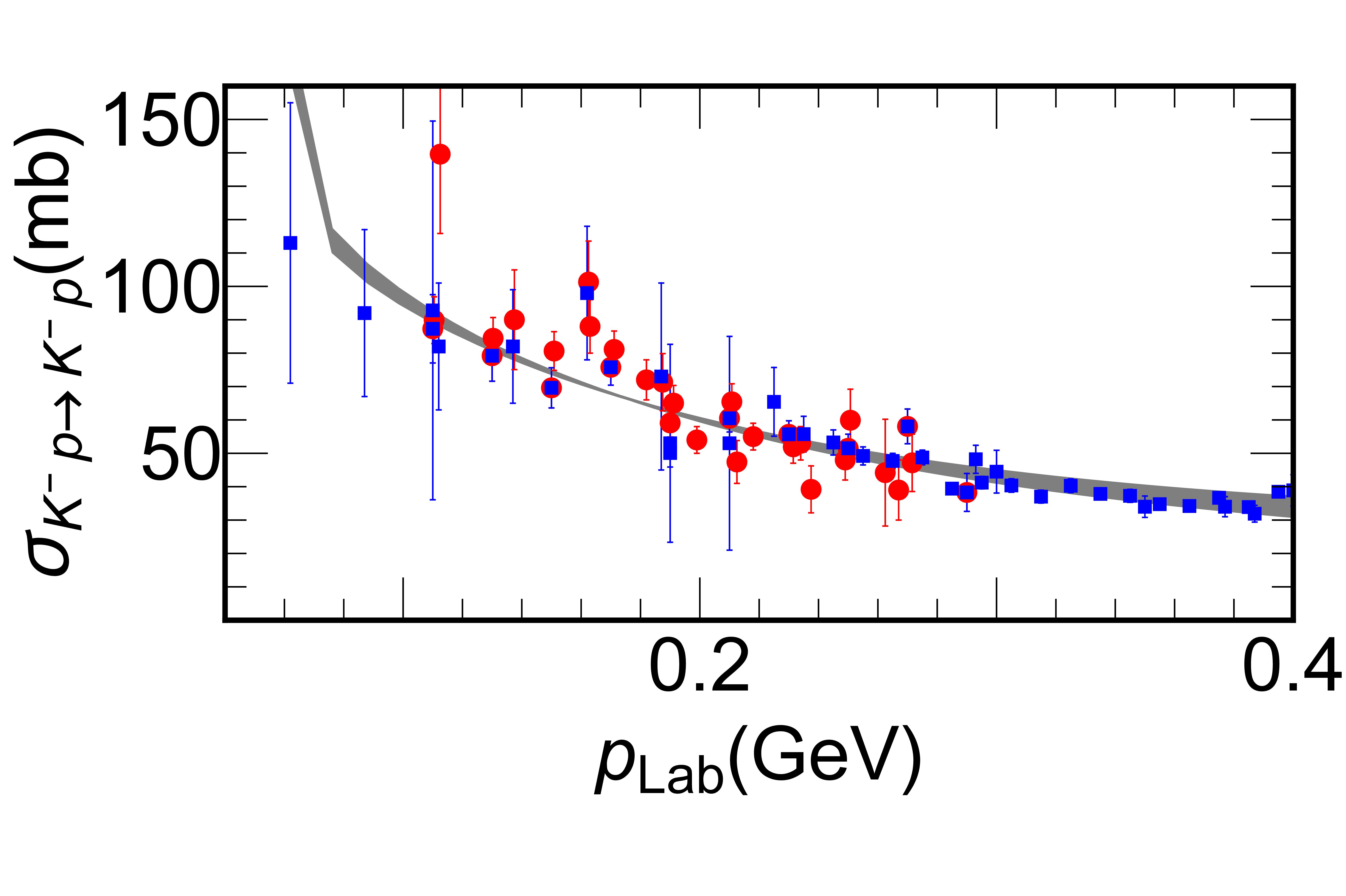}
\caption{Cross sections for  the process $K^- p \to K^- p$. The (red) circles represent the data used in the fitting procedure.  The shaded region represents the results found with Fit I (of Ref.~\cite{Khemchandani:2018amu}).}\label{one}
\end{center}
\end{figure}

We find two sets of solutions with  $\chi^2_\text{d.o.f} \sim 1$ (labelled as Fit I and II in Ref.~\cite{Khemchandani:2018amu}). We show in Fig.~\ref{one} the cross sections, for the $K^-p\to K^-p$ process, obtained from Fit I of Ref.~\cite{Khemchandani:2018amu}. 
Figure~\ref{one} also shows the relevant experimental data for a comparison. The results on the cross sections for other processes obtained within the two types of fits can be found in  Ref.~\cite{Khemchandani:2018amu}. With the two types of fits at hand, we look for the resonances present in the amplitudes by studying them in the complex-energy plane. In the following table 
{\tiny\begin{table}[h!]
{\tablefont Central values of the pole positions, in MeV, of the $(J^P)=(1/2^-)$ states found in each isospin, in the two types of fits found in Ref.~\cite{Khemchandani:2018amu}. }\label{poles01h}
\vspace{0.5cm}
\centering
\begin{tabular}{ccccc}
\hline
&\multicolumn{2}{c}{$\Lambda(1405)$}&\multicolumn{2}{c}{$\Sigma$'s around 1400 MeV}\\
\hline
Fit I&$\makecell{1373-i\,114}$&$\makecell{1426-i\,16}$&$\makecell{1396-i\,5}$&$\makecell{1367-i\,57}$\\
Fit II&$\makecell{1385-i\,124}$&$\makecell{1426-i\,15}$&$-$&$\makecell{1399-i\,36}$\\
\hline
\end{tabular}
\end{table}}
we give the isoscalar and isovector poles present in the s-wave at around 1400 MeV (the uncertainties on the pole positions can be found in Ref.~\cite{Khemchandani:2018amu}). The results obtained in isospin 0 are in good agreement with former works on the topic, including those obtained from the data on photo-/electro-production of $\Lambda(1405)$. As can be seen, we  find poles around the $\bar KN$ threshold in the $I$=1 case also. Two poles appear in Fit I while one pole is found in Fit II. An observation to be mentioned in relation with these $\Sigma$ states is that their masses lie in the energy region where the left hand cut is crossed for some coupled channels. However, we find, interestingly, that  the second pole in Fit I as well as the pole in Fit II continue to appear in the complex plane even if the contribution from the $u$-channel diagrams is switched off.

 \section*{Acknowledgments}
K.P.K. and A.M.T.  gratefully acknowledge the financial support received from Conselho Nacional de Desenvolvimento Cient\'ifico e Tecnol\'ogico (CNPq) (under Grant No. 310759/2016-1 and No. 311524/2016-8). J.A.O. thanks partial financial support from the MINECO (Spain) and FEDER (EU) Grant No. FPA2016-77313-P.  K.P.K also thankfully acknowledges  the support received from the organizers of the conference.

\end{document}